\documentclass[prb,twocolumn,superscriptaddress]{revtex4}
\usepackage{t1enc}
\usepackage{amsmath}
\usepackage{graphicx}
\usepackage{amssymb}
\usepackage{color}

\begin{document}
\providecommand{\eq}{\begin{eqnarray}}
\providecommand{\qe}{\end{eqnarray}}
\providecommand{\e}{\varepsilon}
\providecommand{\h}{\hslash}
\providecommand{\p}{\partial}
\providecommand{\s}{\textrm{sgn}}
\providecommand{\Q}{\widetilde{Q}}
\providecommand{\muTr}{\mu_{\textrm{Tr}}}
\providecommand{\al}{\alpha_{\textsf{l}}}
\providecommand{\aq}{\alpha_{\textsf{qu}}}
\providecommand{\F}{F_i(q,\omega)}
\providecommand{\e}{\varepsilon}
\providecommand{\eF}{\varepsilon_{\textrm{F}}^{}}
\providecommand{\er}{\varepsilon_{\textsc{r}}}
\providecommand{\kf}{k_{\textrm{F}}^{}}
\providecommand{\kfa}{k_{\textrm{F}a}}
\providecommand{\kfb}{k_{\textrm{F}b}}
\providecommand{\vf}{v_{\textrm{F}}^{}}
\providecommand{\vfa}{v_{\textrm{F}a}}
\providecommand{\vfb}{v_{\textrm{F}b}}
\providecommand{\qTF}{q_{\textrm{TF}}^{}}
\providecommand{\s}{\textrm{sign}}
\newcommand{\dd}{\textrm{d}}
\newcommand{\kb}{k_{\textrm{B}}^{}}
\newcommand{\om}{\omega}
\newcommand{\TF}{T_{\textrm{F}}}
\newcommand{\TFa}{T_{\textrm{F}_{a}}}
\newcommand{\TFb}{T_{\textrm{F}_b}}
\newcommand{\acc}{a_{\textsc{c}^{\rule{1mm}{0.2mm}}\textsc{c}}}
\newcommand{\at}{\tilde{a}}
\newcommand{\N}{\mathfrak{n}}
\newcommand{\m}{\mathfrak{m}}
\newcommand{\LL}{\mathfrak{L}}
\newcommand{\rr}{\mathfrak{r}}
\newcommand{\kk}{\mathfrak{K}}
\newcommand{\J}{\mathcal{J}}
\newcommand{\cd}{c^{\dag}}
\newcommand{\cc}{c^{}}
\newcommand{\Iint}{I^{(\textrm{int})}}
\newcommand{\lee}{\ell_{\textrm{e}^{\rule{1mm}{0.2mm}}\textrm{e}}}

\title{Three-particle collisions in quantum wires: \\
  Corrections to thermopower and conductance}
\author{Anders Mathias Lunde}
\affiliation{Nano-Science Center, Niels Bohr Institute,
  University of Copenhagen, DK-2100 Copenhagen, Denmark}
\affiliation{William I. Fine Theoretical Physics Institute, University
  of Minnesota, Minneapolis, MN 55455, USA}

\author{Karsten Flensberg}
\affiliation{Nano-Science Center, Niels Bohr Institute, University
of Copenhagen, DK-2100 Copenhagen, Denmark}

\author{Leonid I. Glazman}
\affiliation{William I. Fine Theoretical Physics Institute,
University of Minnesota, Minneapolis, MN 55455, USA}
\date{\today}

\begin{abstract}
  We consider the effect of electron-electron interaction on the
  electron transport through a finite length single-mode quantum wire with
  reflectionless contacts.  The two-particle scattering events cannot
  alter the electric current and therefore we study the effect of
  three-particle collisions. Within the Boltzmann equation framework,
  we calculate corrections to the thermopower and conductance to the
  leading order in the interaction and in the length of wire $L$. We
  check explicitly that the three-particle collision rate is
  identically zero in the case of several integrable interaction
  potentials. In the general (non-integrable) case, we find a positive
  contribution to the thermopower to leading order in $L$. The
  processes giving rise to the correction involve electron states deep
  in the Fermi sea. Therefore the correction follows an activation law
  with the characteristic energy of the order of the Fermi energy for
  the electrons in the wire.
\end{abstract}

\maketitle

\section{Introduction}

Short clean one-dimensional (1D) mesoscopic wires, often referred to
as quantum point contacts, show conductance
quantization\cite{Wees1988,Wharam1988} as a function of the channel
width. The quantization is well described by the theory of adiabatic
propagation of free electrons.\cite{GlazmanJETPlett1988} For
non-interacting particles, conductance quantization should occur in
longer channels too, as long as there is no backscattering off
inhomogeneities within the channel.

A lot is known about the role of electron-electron interaction of 1D
channels. Electron-electron repulsion in a wire enhances
dramatically the reflection coefficient, making it
energy-dependent.\cite{chang-review1D-2003} However, interaction
between electrons does not alter the quantization (in units of
$2e^2/h$) of an ideal channel conductance in the limit of zero
temperature.\cite{fisherglazman97,Maslov2004} What is still an open
question is whether there are other manifestations of interactions
due to inelastic processes, which influence the transport
properties.

In the absence of interactions, left- and right-moving particles in a
wire are at equilibrium with the reservoirs they originate from.  If a
bias is applied between the reservoirs, then these equilibria differ
from each other, giving rise to a particular form of the
non-equilibrium distribution inside the channel. On the other hand, in
a long ideal channel and in the presence of interactions one may
expect equilibration to occur between the left- and right-movers into
a single distribution characterized by an equilibrium with respect to
a reference frame moving with some drift velocity. Interestingly,
in a model with momentum-independent electron velocity for left- and
right-movers (as it is the case in Tomonaga-Luttinger model) there is
no difference between the two distributions. Effects originating from
the particle-hole asymmetry, however, may discriminate between the
two. Thermopower and Coulomb
drag~\cite{Mortensen2001,Pustilnik-drag-2003,Flensberghcis1996} are
examples of such effects.

\begin{figure}
\centerline{\includegraphics[width=0.4\textwidth,clip=false]{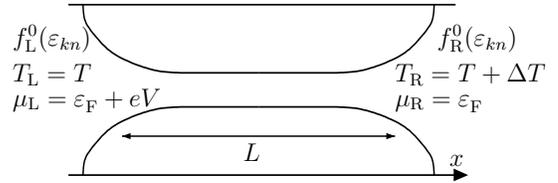}}\caption{A
schematic picture of two metallic gates depleting the underlaying
two dimensional electron gas and thereby forming a short 1D quantum
wire of length $L$. This fabrication method has the advantage of
producing reflectionless contacts to the
leads,\cite{GlazmanJETPlett1988} so that the boundary conditions of
the distribution function is given by the Fermi function of the
reservoirs. We define the thermopower as $S=V/\Delta
T\big|_{I=0}^{}$, i.e.~the voltage $V$ required to counteract a
current due to the temperature difference $\Delta T$.
\label{fig:QPC-picture}}
\end{figure}

At present, little is known about equilibration in a 1D electron
system. In higher dimensions the electron-electron interaction
provides the most effective relaxation mechanism at low temperatures
and therefore we include this relaxation mechanism as the first
approach. However, in 1D pair collisions cannot change the
distribution function for quadratic dispersion, since the momentum
and energy conservation\cite{footnote-pair-scattering-one-band} laws
result in either zero momentum exchange or an interchange of the two
momenta.~\cite{sutherland} In either case the distribution function
remains the same. Thus, the leading equilibration mechanism is due
to three-particle collisions, which we study in this
paper.\cite{khodasCM0702505}

We investigate here the effects of three-particle collisions in
reasonably short wires (see Fig.~\ref{fig:QPC-picture}), where
electron-electron scattering can be considered perturbatively.  As
measurable quantities, we evaluate the temperature dependence of the
thermopower and conductance. Note that for more than one mode
pair collisions become important for certain
fillings.\cite{Lunde-Flensberg-Glazman-2006}

The paper is organized as follows: First we review the
non-interacting limit of thermopower and give a qualitative
explanation of the effects due to three-particle collisions. Then we
describe how to include the electron interactions using the
Boltzmann equation. Next we calculate the main ingredient for our
perturbation theory, namely the three-particle matrix element and
scattering rate using a $T$-matrix expansion. We note several
interesting properties of this scattering rate. Finally, we derive
the conductance and thermopower corrections and discuss the
deviation from the so-called Mott formula. Furthermore, some
technical details are put in two appendices, and in Appendix
\ref{Appendix-A-meso-princip} we show that the number of left and
right movers have to change in a scattering event for the current to
change.

\subsection{Thermopower in the non-interacting limit}

For a wire without interactions the distribution function $f^{(0)}$
is determined solely by the electron reservoirs
\begin{align}
f_{k}^{(0)}=\left\{\begin{array}{ll}
     f^{0}(\e_{k}^{{}}\!-\!\mu_{\textrm{L}}^{{}},T_{\textrm{L}}^{{}})\equiv f^0_{\textrm{L}}(\e_{k}) & \textrm{for}\  k>0,  \\
     f^{0}(\e_{k}^{{}}\!-\!\mu_{\textrm{R}}^{{}},T_{\textrm{R}}^{{}})\equiv f^0_{\textrm{R}}(\e_{k}) & \textrm{for}\ k<0,   \\
\end{array}
\right. \label{eq:f-0}
\end{align}
where $\e_{k}$ is the dispersion relation for momentum $k$ and spin
$\sigma$ (suppressed in the notation), and
$f^{0}(\e-\mu,T)=[1+\exp((\e-\mu)/\kb T)]^{-1}$ is the Fermi
function with $\mu_{\rm L/R}$ and $T_{\rm L/R}$ denoting the
chemical potential and temperature of the left/right contact,
respectively (see Fig.~\ref{fig:QPC-picture}). The electric current
for low temperature $T\ll\TF$ and in linear response to the applied
bias $V$ and temperature difference $\Delta T \ll T$ then follows as
($e>0$)\cite{MolenkampSSAT1992}
\begin{align}
I^{(0)}&=\frac{(-e)}{L}\sum_{\sigma k>0}v_k
\left(f^0_{\textrm{L}}(\e_k)-f^0_{\textrm{R}}(\e_k)\right)\label{eq:I-0}\\
&\simeq -\frac{2e^2}{h}V\left(1-e^{-\TF/T}\right)+\frac{2e}{h}\kb \Delta T
\frac{\TF}{T}e^{-\TF/T}. \label{eq:I-0-sign}
\end{align}
From this the well-known leading-order results for conductance
\begin{equation}
G^{(0)}=\frac{2e^2}{h}\left(1-e^{-\TF/T}\right)
\label{g0}
\end{equation}
and for thermopower
\begin{align}
S^{(0)}=\frac{\kb}{e}\frac{\TF}{T}e^{-\TF/T}
\label{s0}
\end{align}
for a fully open channel are obtained. Here $\TF\equiv\eF/\kb$ is
the Fermi temperature.

\subsection{Main results and a simple picture of the effect of the three-particle scattering}

\begin{figure}
\begin{center}
  \includegraphics[width=0.45\textwidth]{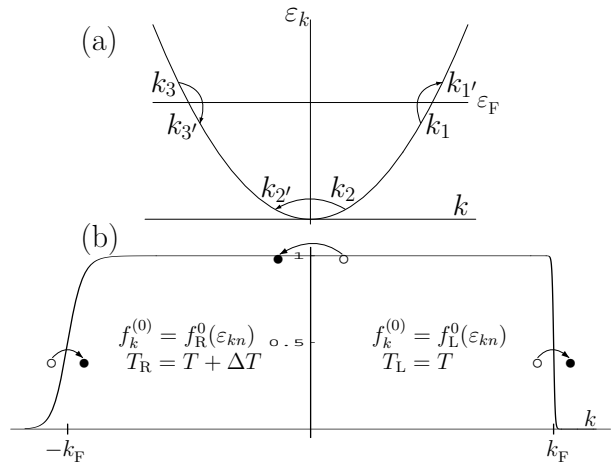}
\caption{(a) The dominant three-particle scattering process at low
temperature in a single energy band.  (b) The three-particle
scattering process perturbing the initial distributions shown with
warm left moving electrons ($k<0$) and cold right moving electrons
($k>0$). Due to the temperature difference of the initial
distributions the scattering process creating left movers dominate
compared to the opposite scattering and therefore it gives a positive
correction to the thermopower. \label{fig:scattering-picture}}
\end{center}
\end{figure}

One of the main results of this paper is that the three-particle
collisions give a \emph{positive} contribution to the thermopower,
i.e.~the current due to a temperature difference is \emph{increased}
by the three-particle scattering. This can be explained in simple
terms. Firstly, to change the current the number of left and right
moving electrons need to change, since it is the number of electrons
going through a mesoscopic structure that determines the current and
not their velocity (see the Appendix \ref{Appendix-A-meso-princip}).
Secondly, we find the dominant scattering process at low temperature
to only involve a single electron changing direction. This occurs
near the bottom of the band as pictured on
Fig.~\ref{fig:scattering-picture}(a). For the initial electronic
distribution the left moving electrons have a higher temperature
then the right moving ones, which favors scattering into the warmer
distribution as seen on Fig.~\ref{fig:scattering-picture}(b). This
thus creates more left moving electrons and thereby increases the
particle current towards the colder reservoir, i.e.~increasing the
thermopower.

Another important point is that the thermopower and conductance
corrections are exponential in temperature, i.e.~proportional to
$\exp(-\TF/T)$. This is a direct consequence of the dominant
three-particle scattering process requiring an empty state near the
bottom of the band. We find the form of the thermopower correction
at low temperatures due to the three particle scattering to be given
by
\begin{equation}
S^{\textrm{int}}\propto L |V|^4 \left(\frac{T}{\TF}\right)^6
\exp(-\TF/T)>0,
\end{equation}
where $V$ is the electron-electron interaction strength and $\TF$
the Fermi temperature. This is found perturbatively in the short
wire limit. The long-wire limit remains an open question,
and we expect that the length dependence of thermopower saturates
once $L$ exceeds some relaxation length (which increases for
decreasing temperature).

In contrast, the conductance correction is negative. To understand
this, note that the chemical potential of the initial distribution
is higher for the right moving electrons then the left moving ones.
This favors scattering into the left moving branch (still with the
process shown in Fig.~\ref{fig:scattering-picture}(a)) for non-zero
temperature and thereby decreasing the current. The form of the
conductance correction is similar to the thermopower correction:
\begin{equation}
G^{\textrm{int}}\propto -L |V|^4 \left(\frac{T}{\TF}\right)^7
\exp(-\TF/T)<0.
\end{equation}

\section{Current calculation in the Boltzmann equation formalism}

\subsection{Effect of interactions on the current}

To model the current through a short 1D quantum wire including
perturbatively the three particle interactions, we use the
Boltzmann equation
\begin{align}
v_{k}\p_{x}f_{k}(x)=\mathcal{I}_{kx}[f],
\label{eq:Boltzmann-eq}
\end{align}
where $f_{k}(x)$ is the distribution function at a space point $x$
between zero and $L$ (see Fig.~\ref{fig:QPC-picture}),
$v_{k}=\frac{1}{\h} \p_k\e_{k}$ is the velocity and
$\mathcal{I}_{kx}[f]$ is the three-body electronic collision
integral, i.e.~no impurity or interface roughness effects are
included here. We include the voltage and temperature difference in
the boundary conditions of the reflectionless
contacts,\cite{GlazmanJETPlett1988} i.e.
\begin{subequations}
\label{eq:Boundary-conditions}
\begin{align}
f_{k}(x=0)&= f^0_{\textrm{L}}(\e_{k})\quad \textrm{for}\quad k>0,\\
f_{k}(x=L)&= f^0_{\textrm{R}}(\e_{k})\quad \textrm{for}\quad k<0,
\end{align}
\end{subequations}
and therefore omit the term $\dot{k}\partial_kf_{k}(x)$ in the
Boltzmann equation, allowed in the linear response
regime.\cite{Sharvin1965}
A similar method has been used to investigate electron-phonon
interactions in short quantum wires,\cite{GurevichPRB1995} quantum
Hall effect in quantum wires \cite{AndoPRB1990} and ballistic
Coulomb drag.\cite{GurevichJPCM1998}

The three-particle collision integral is assumed to be local in
space and is given by
\begin{widetext}
\begin{align}
\mathcal{I}_{k_1x}[f]=&
-\!\!\!\!\!\!\sum_{\substack{\sigma_2\sigma_3\\
\sigma_{1^{\prime}}\sigma_{2^{\prime}}\sigma_{3^{\prime}}}}
\sum_{\substack{k_2k_3\\k_{1^{\prime}}k_{2^{\prime}}k_{3^{\prime}}}}\!\!\!\!
W_{123;1^{\prime}2^{\prime}3^{\prime}}
\left[f_1f_2f_3(1-f_{1^{\prime}})(1-f_{2^{\prime}})(1-f_{3^{\prime}})
-f_{1^{\prime}}f_{2^{\prime}}f_{3^{\prime}}(1-f_1)(1-f_2)(1-f_3)\right],
\label{eq:collision-integral-full}
\end{align}
\end{widetext}
where the quantum numbers are primed/unprimed after/before the
scattering event, $f_i\equiv f_{k_i}(x)$ and the scattering rate
$W_{123;1^{\prime}2^{\prime}3^{\prime}}$ is found in the next
section. Without interactions
($W_{123;1^{\prime}2^{\prime}3^{\prime}}=0$) the solution of the
Boltzmann equation is simply given by $f^{(0)}$ in
Eq.~(\ref{eq:f-0}). When interactions are included it becomes a very
difficult task to solve the Boltzmann equation to all orders in the
interaction. However, for a short wire the interactions only have
short time to change the distribution function away from the initial
distribution $f^{(0)}$ and therefore we expand the distribution
function in orders of $W_{123;1^{\prime}2^{\prime}3^{\prime}}$ as
\begin{align}
f=f^{(0)}+f^{(1)}+\cdots.
\end{align}
To find $f^{(1)}$ to the first order in $W$, we insert the expansion
of $f$ in the Boltzmann equation and realize that only $f^{(0)}$ is
necessary in the collision integral. Since
$\mathcal{I}_{kx}[f^{(0)}]=\mathcal{I}_{k}[f^{(0)}]$ is independent
of $x$, we find that
\begin{subequations}
\label{eq:f-1kk}
\begin{alignat}{2}
f_{k}^{(1)}(x)&=\frac{x}{v_{k}}\mathcal{I}_{k}[f^{(0)}]  \quad & \textrm{for}\quad  &k>0,\\
f_{k}^{(1)}(x)&=\frac{x-L}{v_{k}}\mathcal{I}_{k}[f^{(0)}]  \quad & \textrm{for}\quad &k<0,
\label{eq:f-1}
\end{alignat}
\end{subequations}
using the boundary conditions Eq.~(\ref{eq:Boundary-conditions}).
Therefore the current to the first order in $W$ is
\begin{align}
I=I^{(0)}+e\sum_{\sigma k<0} \mathcal{I}_{k}[f^{(0)}]\equiv
I^{(0)}+I^{\textrm{int}},
\label{eq:I-int-def}
\end{align}
where $I^{(0)}$ is the non-interacting (Landauer) part of the
current from Eq.~(\ref{eq:I-0}) and $I^{\textrm{int}}$ is the part
due to interactions.

\subsection{The linear response limit}\label{sec:linear-response-limit}

The form of the interacting part of the current is now known and the
next step is therefore to evaluate it to linear response to $V$ and
$\Delta T$ to obtain the thermopower and conductance corrections. To
this end, we define $\psi_{k}^{(0)}$ via
\begin{align}
f_{k}^{(0)}\equiv
f^0(\e_{k})+f^0(\e_{k})(1-f^0(\e_{k}))\psi_{k}^{(0)},
\label{eq:psi-0-def}
\end{align}
where $f^0(\e_{k})$ is the Fermi function with temperature $T$ and
Fermi level $\eF$. It turns out that $\psi_{k}^{(0)}$ is
proportional to either $V$ or $\Delta T$. This is seen by using
the identity
\begin{align}
-\kb T\p_{\e}f^0(\e_k)=f^0(\e_{k})(1-f^0(\e_{k})),
\end{align}
so we can identify $\psi_{k}^{(0)}$ by expanding the non-interacting
distribution function $f^{(0)}_k$ (see Eq.~(\ref{eq:f-0}) and
Fig.~\ref{fig:QPC-picture})
\begin{subequations}
\begin{alignat}{2}
f^0_{\textrm{L}}(\e_k)&\simeq f^0(\e_k)+\left[-\p_{\e}f^0(\e_k)\right]eV,\\
f^0_{\textrm{R}}(\e_k)&\simeq
f^0(\e_k)+\left[-\p_{\e}f^0(\e_k)\right](\e-\eF)\frac{\Delta
T}{T},
\end{alignat}
\end{subequations}
i.e.
\begin{align}
\psi_k^{(0)}=\left\{
\begin{array}{ll}
    \frac{eV}{\kb T}, & \hbox{ for $k>0$} \\
    \frac{\e_{k}-\eF}{\kb T}\frac{\Delta T}{T}, & \hbox{ for $k<0$} \\
\end{array}
\right. .
\label{eq:psi-0-value}
\end{align}
Therefore to get $I^{\textrm{int}}$ in linear response to $V$ and
$\Delta T$, we linearized the collision integral
$\mathcal{I}_k[f^{(0)}]$ Eq.~(\ref{eq:collision-integral-full}) with
respect to $\psi_k^{(0)}$ and insert it into $I^{\textrm{int}}$
Eq.(\ref{eq:I-int-def}) to obtain
\begin{align}
I^{\textrm{int}}&=(-e)
\!\!\!\!\!\!\sum_{\substack{\sigma_1\sigma_2\sigma_3\\
\sigma_{1^{\prime}}\sigma_{2^{\prime}}\sigma_{3^{\prime}}}}
\sum_{\substack{k_1<0,k_2k_3\\k_{1^{\prime}}k_{2^{\prime}}k_{3^{\prime}}}}
\Delta_{123;1^{\prime}2^{\prime}3^{\prime}}\nonumber\\
&\times\left[\psi_1^{(0)}+\psi_2^{(0)}+\psi_3^{(0)}-\psi_{1^{\prime}}^{(0)}-\psi_{2^{\prime}}^{(0)}-\psi_{3^{\prime}}^{(0)}\right],
\label{eq:I-1-linear-response}
\end{align}
where we defined
\begin{multline}
\Delta_{123;1^{\prime}2^{\prime}3^{\prime}}=\\
W_{123;1^{\prime}2^{\prime}3^{\prime}}\
f_1^0f_2^0f_3^0(1-f_{1^{\prime}}^0)(1-f_{2^{\prime}}^0)(1-f_{3^{\prime}}^0),
\label{eq:delta-def}
\end{multline}
using the shorthand notation $\psi_i^{(0)}\equiv\psi_{k_i}^{(0)}$
and $f^0_i\equiv f^0 (\e_{k_i})$. To linearize the collision
integral and thereby the correction to the current due to interactions
$I^{\textrm{int}}$, we have used the relation
\begin{multline}
f_1^0f_2^0f_3^0(1-f_{1^{\prime}}^0)(1-f_{2^{\prime}}^0)(1-f_{3^{\prime}}^0)=\\
f_{1^{\prime}}^0f_{2^{\prime}}^0f_{3^{\prime}}^0(1-f_1^0)(1-f_2^0)(1-f_3^0),
\label{eq:H-teorem}
\end{multline}
valid at
$\e_1+\e_2+\e_3=\e_{1^{\prime}}+\e_{2^{\prime}}+\e_{3^{\prime}}$.

Since $\psi_i^{(0)}$ is different for positive and negative $k_i$,
we need to divide the summation in $I^{\textrm{int}}$
Eq.~(\ref{eq:I-1-linear-response}) into positive and negative $k$
sums, which gives $2^5=32$ terms. For this purpose, we introduce the
notation
\begin{align}
\sum_{\substack{k_1<0, k_2>0, k_3<0\\k_{1^{\prime}}>0,
k_{2^{\prime}}>0, k_{3^{\prime}}<0}}
\!\!\!(\cdot)\equiv\sum_{\substack{-+-\\++-}}(\cdot),\quad
\sum_{\substack{\sigma_1\sigma_2\sigma_3\\
\sigma_{1^{\prime}}\sigma_{2^{\prime}}\sigma_{3^{\prime}}}}\!\!\!\!(\cdot)
\equiv \sum_{\textrm{spin}}(\cdot),
\label{eq:notation-def-++}
\end{align}
and similarly for other combinations of the summation intervals. The
32 terms can be simplified to only three terms using energy
conservation and symmetry properties of
$\Delta_{123;1^{\prime}2^{\prime}3^{\prime}}$ in
Eq.~(\ref{eq:delta-def}) under interchange of indices. There are
pairwise exchanges of indices
$\Delta_{123;1^{\prime}2^{\prime}3^{\prime}}=\Delta_{213;1^{\prime}2^{\prime}3^{\prime}}=\Delta_{123;1^{\prime}3^{\prime}2^{\prime}}$
etc. and interchanges between primed and unprimed indices,
$\Delta_{123;1^{\prime}2^{\prime}3^{\prime}}=\Delta_{1^{\prime}2^{\prime}3^{\prime};123}$,
using Eq.~(\ref{eq:H-teorem}) and the fact that
$W_{123;1^{\prime}2^{\prime}3^{\prime}}$ contains a matrix element
squared.  This leads to six terms. Furthermore,
$\Delta_{123;1^{\prime}2^{\prime}3^{\prime}}$ is invariant under
$k_i\rightarrow-k_i$ for all
$i=1,2,3,1^{\prime},2^{\prime},3^{\prime}$ simultaneously due to
time reversal symmetry, also seen explicitly from the form of
$W_{123;1^{\prime}2^{\prime}3^{\prime}}$ (derived below). An example
of how the simplifications occurs can be seen in
Eq.~(\ref{eq:simplification-of-sums}). Thus we obtain the more
compact result
\begin{widetext}
\begin{align}
I^{\textrm{int}}&=
2(-e)\sum_{\textrm{spin}}\sum_{\substack{-++\\+++}}
\Delta_{123;1^{\prime}2^{\prime}3^{\prime}}
\left[\frac{\Delta T}{\kb T^2_{}}(\e_1^{}-\eF)-\frac{eV}{\kb T}\right]\nonumber\\
&+4(-e)\sum_{\textrm{spin}}\sum_{\substack{--+\\+++}}\Delta_{123;1^{\prime}2^{\prime}3^{\prime}}
\left[\frac{\Delta T}{\kb T^2_{}}
\left\{(\e_1-\eF)+(\e_2-\eF)\right\}-\frac{2eV}{\kb
T}\right]\nonumber\\
&+3(-e)\sum_{\textrm{spin}}\sum_{\substack{++-\\+--}}
\Delta_{123;1^{\prime}2^{\prime}3^{\prime}} \left[\frac{\Delta
T}{\kb T^2_{}}
\left\{-(\e_3-\eF)+(\e_{2^{\prime}}-\eF)+(\e_{3^{\prime}}-\eF)\right\}-\frac{eV}{\kb
T}\right], \label{eq:I-1-linear-simplificeret-exact}
\end{align}
\end{widetext}
where the definition of $\psi_i^{(0)}$ in Eq.~(\ref{eq:psi-0-value})
was inserted. An important point is that the number of
positive/negative wave-vector intervals is not the same before and
after the scattering. Therefore, we note that only scattering events
that change the number of left and  right moving electrons
contributes to the interaction correction to the current. The origin
of this is the cancellation of the velocity in the definition of the
current and in the distribution functions \eqref{eq:f-1kk}.

This cancellation thus leads to an expression for the interaction
correction to the current in
Eq.~\eqref{eq:I-1-linear-simplificeret-exact} where all the in-going
and out-going momenta enter on equal footing. In the Appendix
\ref{Appendix-A-meso-princip}, we show that this is valid to all
orders in perturbation theory. Due to this property and momentum
conservation, there are no processes that alter the current possible
near the Fermi level. Consequently, states far away from the Fermi
level have to be involved in the scattering, which, as we will see,
leads to a suppression of $I^{\textrm{int}}$ by a factor
$\exp(-\TF/T)$. The distribution function on the other hand,
\emph{can} be changed by scattering processes near the Fermi level.

To identify the important processes we find in the next section the
scattering rate $W_{123;1^{\prime}2^{\prime}3^{\prime}}$.

\begin{figure*}
\centerline{\includegraphics[width=0.97\textwidth,clip=false]{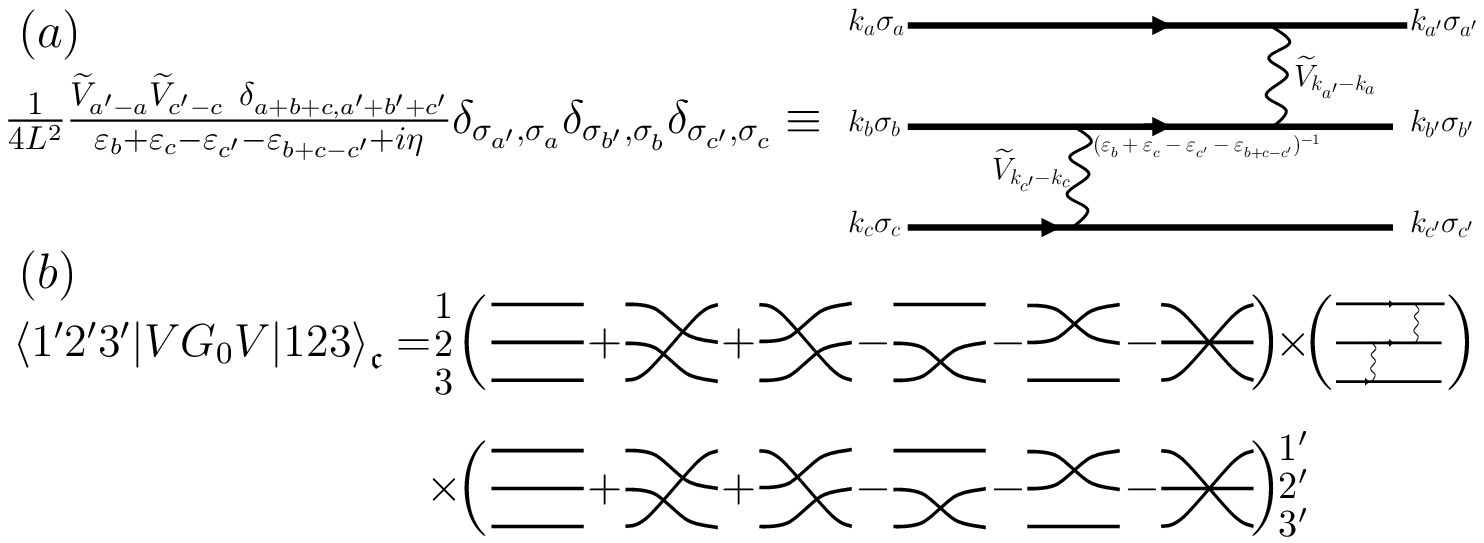}}
\caption{A visualization of the connected three-particle scattering
matrix element Eq.~(\ref{eq:matrix-element-VGV-genneral}), where
three particles interchange their momenta and energy. This matrix
element enters the scattering rate via the generalized Fermi Golden
rule Eq.~(\ref{eq:W-fermi-rule}). (a) The basic three-particle
interaction consisting of two interaction lines and a free
propagation, see Eq.~(\ref{eq:matrix-element-VGV-genneral}). (b)
Picture of the exchange processes times the basic interaction needed
to form the matrix element
Eq.~(\ref{eq:matrix-element-VGV-genneral}).
\label{fig:diagram}}
\end{figure*}

\section{Three-particle scattering rate}

The three particle scattering rate
$W_{123;1^{\prime}2^{\prime}3^{\prime}}$ is calculated using the
generalized Fermi Golden rule inserting the $T$-matrix, $T\equiv
V+VG_0T$, iterated to second order in the interaction $V$ to get the
three particle interaction amplitude, i.e.,
\begin{align}
W_{123;1^{\prime}2^{\prime}3^{\prime}}=\frac{2\pi}{\h}|\langle
1^{\prime}2^{\prime}3^{\prime}|VG_0V|123\rangle_{\mathfrak{c}}|^2\delta(E_i-E_f),
\label{eq:W-fermi-rule}
\end{align}
where $E_i^{}=\e_1^{}+\e_2^{}+\e_3^{}$ is the initial energy,
$E_f^{}=\e_{1^{\prime}}+\e_{2^{\prime}}+\e_{3^{\prime}}$ the final
energy, $G_0$ is the resolvent operator (or free Green function),
$j$ is short hand for $k_j$, and the subscript '$\mathfrak{c}$'
means connected in the sense that the scattering process cannot be
effectively a two particle process, where one of the incoming
particles does not participate in the scattering. Explicitly
$G_0$ and $V$ are given by
\begin{align}
G_0&=\frac{1}{E_i-H_0+i\eta}, \qquad (\eta\rightarrow 0^+)\\
V&=\frac{1}{2L}\sum_{k_1^{}k_2^{}q}\sum_{\sigma_1^{}\sigma_2^{}}V_q
c_{k_1^{}+q\sigma_{1}^{}}^{\dag} c_{k_2^{}-q\sigma_{2}^{}}^{\dag}
c_{k_2^{}\sigma_{2}^{}}^{} c_{k_1^{}\sigma_{1}^{}}^{}.
\end{align}
Here $H_0$ is the unperturbed Hamiltonian (i.e.~kinetic energy with
some dispersion), $V_q$ the Fourier-transformed interaction potential,
and $c_{k\sigma}^{}$ ($c_{k\sigma}^{\dag}$) is the annihilation
(creation) operator. To calculate the matrix element $\langle
1^{\prime}2^{\prime}3^{\prime}|VG_0V|123\rangle_{\mathfrak{c}}$, we
write the initial and final states as
\begin{align}
|123\rangle&=c_{k_1^{}\sigma_{1}^{}}^{\dag} c_{k_2^{}\sigma_{2}^{}}^{\dag}c_{k_3^{}\sigma_{3}^{}}^{\dag}|0\rangle,\\
|1^{\prime}2^{\prime}3^{\prime}\rangle&=
c_{k_{1^{\prime}}^{}\sigma_{1^{\prime}}^{}}^{\dag}
c_{k_{2^{\prime}}^{}\sigma_{2^{\prime}}^{}}^{\dag}
c_{k_{3^{\prime}}^{}\sigma_{3^{\prime}}^{}}^{\dag}|0\rangle,
\end{align}
where $|0\rangle$ is the empty state. Using the anti-commutator
algebra $\{c_i^{},c_j^{\dag}\}=\delta_{i,j}$, we obtain
\begin{align}
G_0V|123\rangle=&\frac{1}{2L}\sum_q V_q\label{eq:VG|i>}\\
&\times\!\!\!\!\!\!\!\!\!\!\sum_{(abc)\in P(123)}
\!\!\frac{\s(abc)}{\e_{ba}(q)} c_{k_a^{}+q\sigma_{a}^{}}^{\dag}
c_{k_b^{}-q\sigma_{b}^{}}^{\dag}
c_{k_c^{}\sigma_{c}^{}}^{\dag}|0\rangle, \nonumber
\end{align}
where we introduced
\begin{align}
\e_{ba}(q)&=\e_b+\e_a-\e_{b-q}-\e_{a+q}+i\eta\nonumber\\&=\frac{\h^2}{m}q(k_b-k_a-q)+i\eta
\end{align}
(the last equality is only valid for a quadratic dispersion), and
where the set of permutations is given by
$P(123)=\{(123)^+,(231)^+,(312)^+,(132)^-,(321)^-,(213)^-\}$. Here
the signs of the permutation, $\s(abc)$,  are shown as superscripts.

In order to exclude the effectively two-particle processes when
multiplying Eq.~(\ref{eq:VG|i>}) by $\langle
1^{\prime}2^{\prime}3^{\prime}|V$ from the left, $k_c^{}$
($c=1,2,3$) needs to be different from $k_{j^{\prime}}$ ($j=1,2,3$).
The result is
\begin{widetext}
\begin{align}
\langle
1^{\prime}2^{\prime}3^{\prime}|VG_0V|123\rangle_{\mathfrak{c}}&=\frac{1}{(2L)^2}\!\!\!\!\!
\sum_{(abc)\in P(123)} \sum_{(a^{\prime}b^{\prime}c^{\prime})\in
P(1^{\prime}2^{\prime}3^{\prime})}\!\!\!\!\!\!\!\!\!\!\!\!\!
\s(abc)\s(a^{\prime}b^{\prime}c^{\prime})
\frac{\widetilde{V}_{a^{\prime}-a}\widetilde{V}_{c^{\prime}-c} \
\delta_{a+b+c,a^{\prime}+b^{\prime}+c^{\prime}}}{\e_b+\e_c-\e_{c^{\prime}}-\e_{b+c-c^{\prime}}+i\eta}
\delta_{\sigma_{a^{\prime}},\sigma_a^{}}
\delta_{\sigma_{b^{\prime}},\sigma_b^{}}
\delta_{\sigma_{c^{\prime}},\sigma_c^{}},
\label{eq:matrix-element-VGV-genneral}
\end{align}
where $\widetilde{V}_q=V_q+V_{-q}$ is the symmetrized interaction.
The matrix element consist of 36 terms and the scattering rate thus
has $36^2=1296$ terms. To obtain this result we did not use energy
conservation. For a quadratic dispersion, the denominator is only
zero if we have an effective pair collision or if the momentum
transfer is zero, as seen from the expression
$\e_{ba}(q)=\frac{\h^2}{m}q(k_b-k_a-q)+i\eta$. A picture of the
matrix element is found in Fig.~\ref{fig:diagram}(b), where the
exchange processes (including the sign) are visualized as different
ways to connect two interaction lines and an intermediate
propagation ($G_0$) seen on Fig.~\ref{fig:diagram}(a). The inclusion
of the Fermi statistics makes a substantial difference for the
properties of the scattering rate as compared to the case described
in Ref.~[\onlinecite{Vasilopoulos1994}], which is obtained by
setting all $\s(\cdots)=+1$.

We can rewrite the matrix element
Eq.~(\ref{eq:matrix-element-VGV-genneral}) in a more transparent way
in terms of quantum mechanical exchange symmetry. First, we
introduce the following combination of three-particle scattering
amplitudes
\begin{align}
\mathbb{V}(11^{\prime},22^{\prime},33^{\prime})=&
\frac{\delta_{\sigma_{1^{\prime}},\sigma_1^{}}
\delta_{\sigma_{2^{\prime}},\sigma_2^{}}
\delta_{\sigma_{3^{\prime}},\sigma_3^{}}}{4L^2}
\bigg[\frac{\widetilde{V}_{1^{\prime}-1}\widetilde{V}_{3^{\prime}-3}}{\e_3+\e_2-\e_{3^{\prime}}-\e_{2+3-3^{\prime}}}+
\frac{\widetilde{V}_{2^{\prime}-2}\widetilde{V}_{1^{\prime}-1}}{\e_1+\e_3-\e_{1^{\prime}}-\e_{3+1-1^{\prime}}}
+\frac{\widetilde{V}_{3^{\prime}-3}\widetilde{V}_{2^{\prime}-2}}{\e_2+\e_1-\e_{2^{\prime}}-\e_{1+2-2^{\prime}}}\nonumber\\
&+\frac{\widetilde{V}_{1^{\prime}-1}\widetilde{V}_{2^{\prime}-2}}{\e_2+\e_3-\e_{2^{\prime}}-\e_{3+2-2^{\prime}}}+
\frac{\widetilde{V}_{3^{\prime}-3}\widetilde{V}_{1^{\prime}-1}}{\e_1+\e_2-\e_{1^{\prime}}-\e_{2+1-1^{\prime}}}+
\frac{\widetilde{V}_{2^{\prime}-2}\widetilde{V}_{3^{\prime}-3}}{\e_3+\e_1-\e_{3^{\prime}}-\e_{1+3-3^{\prime}}}\bigg],
\label{eq:V-def}
\end{align}
\end{widetext}
and after some rewriting, we then obtain
\begin{align}
\langle
1^{\prime}2^{\prime}3^{\prime}&|VG_0V|123\rangle_{\mathfrak{c}}=
\delta_{k_1+k_2+k_3,k_{1^{\prime}}+k_{2^{\prime}}+k_{3^{\prime}}}\label{eq:matrix-element-vha-V}\\
\Big[&\mathbb{V}(11^{\prime},22^{\prime},33^{\prime})
+\mathbb{V}(12^{\prime},23^{\prime},31^{\prime})
+\mathbb{V}(13^{\prime},21^{\prime},32^{\prime})\nonumber\\
-&\mathbb{V}(11^{\prime},23^{\prime},32^{\prime})
-\mathbb{V}(13^{\prime},22^{\prime},31^{\prime})
-\mathbb{V}(12^{\prime},21^{\prime},33^{\prime})\Big].\nonumber
\end{align}
We interpret this result in a way similar to a two-particle matrix
element,
\begin{align}
\langle & 1^{\prime}2^{\prime}|V|12\rangle=
\frac{\delta_{k_{1}^{}+k_{2}^{},k_{1^{\prime}}^{}+k_{2^{\prime}}^{}}
}{L}\nonumber\\
&\times\left[V_{k_{1^{\prime}}^{}-k_1^{}}^{}
\delta_{\sigma_{1^{}}^{},\sigma_{1^{\prime}}}\delta_{\sigma_{2^{}}^{},\sigma_{2^{\prime}}}
- V_{k_{2^{\prime}}-k_{1^{}}^{}}^{}
\delta_{\sigma_{1^{}}^{},\sigma_{2^{\prime}}}\delta_{\sigma_{2^{}}^{},\sigma_{1^{\prime}}}
\right],
\label{eq:2-particle-matrix-element}
\end{align}
which contains a direct (first term) and a exchange term (where
$1^{\prime}\leftrightarrow2^{\prime}$).

In the three-particle case
$\mathbb{V}(11^{\prime},22^{\prime},33^{\prime})$ is the direct term
and one can make five exchange processes (instead of one) by
exchanging the three final states $1^{\prime}$, $2^{\prime}$ and
$3^{\prime}$. This gives Eq.~(\ref{eq:matrix-element-vha-V}). The
sign in front of each $\mathbb{V}(\cdots)$ is determined by the
number of exchanges made, e.g.~in
$\mathbb{V}(11^{\prime},23^{\prime},32^{\prime})$ a single exchange,
$2^{\prime}\leftrightarrow3^{\prime}$, gives a minus $(-1)^1$
whereas for $\mathbb{V}(12^{\prime},23^{\prime},31^{\prime})$ two
exchanges ($1^{\prime}\leftrightarrow3^{\prime}$ followed by
$3^{\prime}\leftrightarrow2^{\prime}$) gives a positive sign
$(-1)^2$. Furthermore, the arguments in
$\mathbb{V}(11^{\prime},22^{\prime},33^{\prime})$ are ordered in
three pairs such that the differences between the elements in each
pair are the only arguments of the interaction potential, see
Eq.~(\ref{eq:V-def}). This is useful when constructing
approximations having a specific scattering process in mind.

How the matrix element was rewritten into the form of
Eq.~(\ref{eq:matrix-element-vha-V}) can also be described in terms of
the drawings of Fig.~\ref{fig:diagram}. The direct term
$\mathbb{V}(11^{\prime},22^{\prime},33^{\prime})$ is the sum of the
six terms having mirror-symmetric exchanges before and after the
scattering.  The other terms in
Eq.(\ref{eq:matrix-element-vha-V}) then can be obtained by suitable
changes of outgoing lines.

\subsection{Zero three-particle scattering rate for integrable models}
\label{subsec:point-like-int}

The expressions we obtain for the three-particle scattering rates
Eq.~(\ref{eq:W-fermi-rule}) are quite cumbersome. Nevertheless, the
obtained results allow for some consistency checks. Remarkably, for
some two-body potentials, scattering of the particles of an $N$-body
system is exactly equivalent to a sequence of two-body collisions.
Such ``special'' potentials were studied in the context of
integrable quantum many-body problems.\cite{sutherland} We recall
now that for a quadratic band a pair collision does not change the
momenta of the incoming particles or simply permutes the two
momenta. Therefore, three-particle scattering for a the integrable
potentials may result {\it only} in permutations within the group of
three momenta of the colliding particles; all other three-particle
scattering amplitudes must be zero for such potentials. In the
context of this work it means that even three-particle (or
higher-order) collisions would not bring electron equilibration for
such types of electron-electron interaction.

In this section we check that the three-particle scattering amplitudes
are indeed zero for two special potentials.

\subsubsection{Point-like interaction}

In the case of contact interaction,
$\widetilde{V}_q=\textrm{constant}\equiv\widetilde{V}_0$, and for
any kind of electron dispersion relation (i.e., not necessarily
quadratic), we find by using the energy conservation law that
\begin{align}
\sum_{\textrm{spin}}|\langle
1^{\prime}2^{\prime}3^{\prime}|&VG_0V|123\rangle_c|^2
=\frac{2\widetilde{V}_0^4}{(2L)^4}\delta_{k_1+k_2+k_3,k_{1^{\prime}}+k_{2^{\prime}}+k_{3^{\prime}}}
\nonumber\\
\times
\Big[&|A_{121^{\prime}}-A_{122^{\prime}}-A_{131^{\prime}}+A_{132^{\prime}}|^2\nonumber\\
&+|A_{121^{\prime}}-A_{123^{\prime}}-A_{131^{\prime}}+A_{133^{\prime}}|^2\nonumber\\
&+|A_{122^{\prime}}-A_{123^{\prime}}-A_{132^{\prime}}+A_{133^{\prime}}|^2\nonumber\\
&+|A_{121^{\prime}}-A_{122^{\prime}}-A_{231^{\prime}}+A_{232^{\prime}}|^2\nonumber\\
&+|A_{131^{\prime}}-A_{132^{\prime}}-A_{231^{\prime}}+A_{232^{\prime}}|^2\nonumber\\
&+|A_{121^{\prime}}-A_{123^{\prime}}-A_{231^{\prime}}+A_{233^{\prime}}|^2\nonumber\\
&+|A_{131^{\prime}}-A_{133^{\prime}}-A_{231^{\prime}}+A_{233^{\prime}}|^2\nonumber\\
&+|A_{122^{\prime}}-A_{123^{\prime}}-A_{232^{\prime}}+A_{233^{\prime}}|^2\nonumber\\
&+|A_{132^{\prime}}-A_{133^{\prime}}-A_{232^{\prime}}+A_{233^{\prime}}|^2\Big],
\label{eq:V-const-E-general}
\end{align}
where $A_{abc}=(\e_a+\e_b-\e_c-\e_{a+b-c}+i\eta)^{-1}$. This is a
major simplification from $36^2=1296$ to $9\times 4^2=144$ terms by
performing the spin summation. If furthermore, the dispersion is
quadratic, $\e_k\propto k^2$, then we find the (at first sight)
surprising cancellation
\begin{align}
\sum_{\textrm{spin}}|\langle
1^{\prime}2^{\prime}3^{\prime}&|VG_0V|123\rangle_c|^2\nonumber\\
&\times\delta(\e_1+\e_2+\e_3-\e_{1^{\prime}}-\e_{2^{\prime}}-\e_{3^{\prime}})=0.
\end{align}
This can be seen directly from Eq.(\ref{eq:V-const-E-general}) or
by noting that
\begin{align}
\!\mathbb{V}(1a^{\prime},2b^{\prime},3c^{\prime})\delta(\e_1+\e_2+\e_3-\e_{a^{\prime}}-\e_{b^{\prime}}-\e_{c^{\prime}})=0,
\end{align}
for a quadratic dispersion and constant interaction for
$(a^{\prime}b^{\prime}c^{\prime})\in
P(1^{\prime}2^{\prime}3^{\prime})$, i.e.~each term of
Eq.~(\ref{eq:matrix-element-vha-V}) is zero.
$\mathbb{V}(11^{\prime},22^{\prime},33^{\prime})$ cancels in such
a way, that the three first terms of Eq.~(\ref{eq:V-def}) cancel
each other (the even permutations of $(123)$ combined with the
same primed permutation) and the three last terms cancel each
other (the odd permutations of $(123)$ combined with the same
primed permutation).

In fact, the described above cancellation is in agreement with the
general factorization results for the $S$--matrix of one-dimensional
$N$--body problem with $\delta$--function interaction in real
space.\cite{yang} In this context, it is crucial that the particles
have quadratic dispersion relation; if we use e.g.~$\e_k\propto
k^4$, then the cancellation does not occur. Notice also that the
cancellation we demonstrate is not a trivial zero. Indeed, the
underlying two-particle amplitudes
Eq.~(\ref{eq:2-particle-matrix-element}) are finite for a
$q$--independent potential, if one includes spins. (For spinless
fermions and contact interaction the matrix element would be zero
because the direct and the exchange terms cancel in accordance with
the Pauli principle.)

\subsubsection{The $\widetilde{V}_q=V_0(1-q^2/q_0^2)$ interaction}

We checked also that the energy conserving part of the matrix
element $\langle
1^{\prime}2^{\prime}3^{\prime}|VG_0V|123\rangle_{\mathfrak{c}}$ in
the case of {\it spinless} fermions, quadratic dispersion, and the
Fourier transformed interaction potential of the form
\begin{equation}
\widetilde{V}_q=V_0\left(1-\frac{q^2}{q_0^2}\right),
\label{liebliniger}
\end{equation}
becomes equal to zero. This is also possible to expect because of
the relation of the potential Eq.~(\ref{liebliniger}) to the
integrable 1D bosonic Lieb-Liniger model~\cite{liebliniger}. Indeed,
the bosonic model with contact interaction potential $\propto
g_B\delta (x_1-x_2)$ may be exactly mapped~\cite{cazalilla} onto the
spinless fermionic model with interaction $V_F(x_1-x_2)\propto
-(1/g_B)\delta^{\prime\prime} (x_1-x_2)$. The integrability of the
bosonic model guarantees the integrability of the corresponding
Fermionic one. Adding a contact interaction to $V_F$ does no harm,
as we are considering spinless Fermions. Finally, Fourier
transformation takes us to Eq.~(\ref{liebliniger}).

We checked that including the spin degree of freedom, spoils the
remarkable cancellation for a three-particle amplitude.

In the following Sections, we assume a general case interaction
potential for which the three-particle scattering amplitudes lead to
a non-trivial re-distribution of the momenta between the particles.

\section{Thermopower and conductance corrections due to three particle interaction}

In this section, we go though the main ideas and approximations in
evaluating the current correction due to interactions
$I^{\textrm{int}}$ Eq.~(\ref{eq:I-1-linear-simplificeret-exact})
to lowest order in the temperature, $T\ll\TF$. We give a more
detailed calculation in the Appendix
\ref{Appendix-B-detailed-calcualtion}.

As noted previously, all three terms in $I^{\textrm{int}}$
Eq.~(\ref{eq:I-1-linear-simplificeret-exact}) are exponentially
suppressed, since momentum conservation
$$k_{1^{}}+k_{2^{}}+k_{3^{}}=k_{1^{\prime}}+k_{2^{\prime}}+k_{3^{\prime}}$$
forbids scattering processes near the Fermi level for the given
combinations of positive and negative $k$ intervals. To be more
specific, it is the phase space restrictions of the Fermi functions
that give the exponential suppression, i.e.
\begin{align}
f_1^0f_2^0f_3^0(1-f_{1^{\prime}}^0)(1-f_{2^{\prime}}^0)(1-f_{3^{\prime}}^0)\propto
e^{-\TF/T}.
\end{align}
We begin by identifying the most important three particle scattering
process. The three terms in $I^{\textrm{int}}$
Eq.~(\ref{eq:I-1-linear-simplificeret-exact}) are: (i) two right
movers backscattering a left mover while remaining right movers; (ii)
one right mover keeping its direction, while backscattering two left
movers, and (iii) a left and a right mover keeping their directions,
while backscattering the third particle. From now on, we will
concentrate on the case of Coulomb interaction $\widetilde{V}_q$,
which is the largest for small $q$, therefore we want to identify
processes where the initial and final state are close in momentum
space~\cite{coulomb}. Further, the process(es) should not require more than one
electron in states suppressed exponentially by the Fermi functions.
One can see that due to the constraints stemming from momentum and
energy conservation, in fact only process (iii) allows both initial
and final states to be close to each other in momentum space and at
the same time having only a single exponentially suppressed factor.
The corresponding scattering process is of the type shown in
Fig.~\ref{fig:scattering-picture}(a).  Therefore to the first order in
$\exp(-\TF/T)$, we include only the third one in
Eq.~(\ref{eq:I-1-linear-simplificeret-exact}). This leads to
\begin{align}
I^{\textrm{int}}\simeq&\
3(-e)\sum_{\textrm{spin}}\sum_{\substack{++-\\+--}}
\Delta_{123;1^{\prime}2^{\prime}3^{\prime}}\nonumber\\
&\times\left[\frac{\Delta T}{\kb T^2_{}}
\left\{-\e_3+\e_{2^{\prime}}+\e_{3^{\prime}}-\eF\right\}-\frac{eV}{\kb
T}\right].
\label{eq:I-int-vigtigt}
\end{align}
Here $\Delta_{123;1^{\prime}2^{\prime}3^{\prime}}$ expresses the
available phase space in form of the Fermi functions and the
three-particle scattering rate, see Eq.~(\ref{eq:delta-def}).

One essential approximation is that for the scattering process
depicted in Fig.~\ref{fig:scattering-picture}(a), we may replace the
full Fermi distribution functions by the exponential tales or the
low-temperature limit expressions, i.e.
\begin{subequations}
\label{f0s}
\begin{alignat}{2}
f^0_1&\simeq \theta(\kf-k_1)\theta(k_1), & \quad
1-f^0_{1^{\prime}}&\simeq \theta(k_{1^{\prime}}-\kf),\label{eq:Fermi-function-1-1m}\\
f^0_2&\simeq \theta(\kf-k_2)\theta(k_2), & \quad
1-f^0_{2^{\prime}}&\simeq e^{(\e_{2^{\prime}}-\eF)/\kb T}, \label{eq:Fermi-function-2-2m}\\
f^0_3&\simeq e^{-(\e_{3^{}}-\eF)/\kb T}, & \quad
1-f^0_{3^{\prime}}&\simeq e^{(\e_{3^{\prime}}-\eF)/\kb
T}\label{eq:Fermi-function-3-3m}.
\end{alignat}
\end{subequations}
Note that $k_1$, $k_{1^{\prime}}$ and $k_2$ are all positive. We see
that the product of the Fermi functions is indeed exponentially
suppressed, i.e.~$\propto \exp(-\TF/T)$.

The second essential approximation is that for the scattering process
seen in Fig.~\ref{fig:scattering-picture}(a) the initial and final
states differ by a small momentum. Therefore the matrix element in the
transition rate $W_{123;1^{\prime}2^{\prime}3^{\prime}}$ is dominated
by the direct term $\mathbb{V}(11^{\prime},22^{\prime},33^{\prime})$
in Eq.~(\ref{eq:matrix-element-vha-V}), since the five exchange terms
are suppressed by the Coulomb interaction $|\widetilde{V}_{|q|\sim
  \kf}|\ll|\widetilde{V}_{|q|\ll \kf}|$, i.e.
\begin{multline}
\langle
1^{\prime}2^{\prime}3^{\prime}|VG_0V|123\rangle_{\mathfrak{c}}\simeq \\
\delta_{k_1+k_2+k_3,k_{1^{\prime}}+k_{2^{\prime}}+k_{3^{\prime}}}
\mathbb{V}(11^{\prime},22^{\prime},33^{\prime}).
\label{eq:matrix-element-direct-only}
\end{multline}

The direct term Eq.~(\ref{eq:matrix-element-direct-only}) would be
zero for $\widetilde{V}_{q}=\textrm{const}$. In the case of
quadratic dispersion relation and general $\widetilde{V}_{q}$, it
vanishes in the limit $(k_{i^{\prime}}-k_i)\to 0$ and
$(k_{j^{\prime}}-k_j)\rightarrow 0$ for $i,j\in \{1,2,3\}$  due to
the Pauli principle. For a quadratic dispersion and for a general
symmetrized interaction $\widetilde{V}_{q}=V_q+V_{-q}$, the direct
term $\mathbb{V}(11^{\prime},22^{\prime},33^{\prime})$ simplifies to
the following expression
\begin{align}
\mathbb{V}&(11^{\prime},22^{\prime},33^{\prime})=
\frac{\delta_{\sigma_{1^{\prime}},\sigma_1^{}}
\delta_{\sigma_{2^{\prime}},\sigma_2^{}}
\delta_{\sigma_{3^{\prime}},\sigma_3^{}}}{4L^2 \h^2/m}(q_1+q_3) \nonumber\\
&\times\frac{\left[-(q_1+q_3)\widetilde{V}_{q_1}\widetilde{V}_{q_3}+\widetilde{V}_{q_1+q_3}(q_3\widetilde{V}_{q_1}+q_1\widetilde{V}_{q_3})\right]}{(k_1-k_3+q_1)q_1q_3(k_1-k_3-q_3)}
\label{eq:V-direct-quadratic-sec4}
\end{align}
where we used energy conservation and introduced
$q_1=k_{1^{\prime}}-k_1$ and $q_3=k_{3^{\prime}}-k_3$.

Next, we give a qualitative explanation for the power law in $T$ for
the interacting current correction Eq.~(\ref{eq:I-int-vigtigt})
using the quadratic dispersion. First, we consider the phase space
constraint. To do the sum over all $k$ in
Eq.~(\ref{eq:I-int-vigtigt}) we use the momentum and energy
conservation and introduce new variables $q_1=k_{1^{\prime}}-k_1$
and $q_3=k_{3^{\prime}}-k_3$, i.e., change the summation variables,
\begin{align}
k_1,k_2,k_3,k_{1^{\prime}},k_{2^{\prime}},k_{3^{\prime}} \to
k_1,k_3,q_1,q_3.
\end{align}
The energy conservation for a quadratic dispersion gives a factor of
$1/|q_1+q_3|$ (see e.g.~Eq.~(\ref{eq:energy-conservation})). For the
process at hand, the $k_1$ and $k_3$ are close to the Fermi level
and each of their sums contribute with a factor of $q_1$ and $q_3$,
respectively. The Fermi functions give the exponential suppression
and a contribution to the phase space in form of an exponential
tail, i.e.
\begin{multline}
f^0_1f^0_2f^0_3(1-f^0_{1^{\prime}})(1-f^0_{2^{\prime}})(1-f^0_{3^{\prime}})\propto\\
e^{-\TF/T}e^{(\e_{2^{\prime}}-\e_3+\e_{3^{\prime}})/\kb T}
\end{multline}
see Eq.~(\ref{eq:Fermi-function-1-1m}-\ref{eq:Fermi-function-3-3m}).
To get the low temperature result for $I^{\textrm{int}}$,
Eq.~(\ref{eq:I-int-vigtigt}), we use the method of steepest decent
to calculate the integral. To this end, we note that the exponent
$\e_{2^{\prime}}-\e_3+\e_{3^{\prime}}$ is a function of $q_1$ and
$q_3$ and in the limit $T/\TF\to 0$ the most important part is
around the origin $q_1=q_3=0$. Here
$\e_{2^{\prime}}-\e_3+\e_{3^{\prime}}$ vanishes as $-\frac{1}{2}\h
\vf(q_1+q_2)$ (see Appendix \ref{Appendix-B-detailed-calcualtion}
for details). Therefore collecting the phase space factors the
current correction due to three particle interactions
Eq.~(\ref{eq:I-int-vigtigt}) becomes
\begin{align}
I^{\textrm{int}}\propto\
&\e^{-\TF/T} \int \! \dd q_1\int \! \dd q_3
\frac{q_1q_3}{|q_1+q_3|} e^{-\frac{\TF}{T}(q_1+q_3)/\kf}
\nonumber\\
&\times|\mathbb{V}(11^{\prime},22^{\prime},33^{\prime})|^2\nonumber\\
&\times\left[\frac{\Delta T}{T}\frac{\TF}{T}
\left(-\frac{q_1+q_3}{\kf}-1\right)-\frac{eV}{\kb T}\right].
\end{align}
in the limit $T\ll\TF$. Furthermore, it turns out that the constraints
$k_2>0$ and $k_{2^{\prime}}<0$ in the sum Eq.~(\ref{eq:I-int-vigtigt})
only leaves phase space close to $q_1=q_3$ for $T/\TF\to 0$, so we can
set $q_3=q_1$ in the integrand and do the integral over $q_3$, which
is $\propto q_1^2$ due to the phase space limits. To lowest order in
temperature this yields
\begin{align}
I^{\textrm{int}}\propto\
\e^{-\TF/T} \int_0^{\infty} \!\!\! \dd q& \ q^3
e^{-\frac{\TF}{T}2q/\kf}
|\mathbb{V}(11^{\prime},22^{\prime},33^{\prime})|^2\nonumber\\
&\times\left[\frac{\Delta T}{T}\frac{\TF}{T}+\frac{eV}{\kb
T}\right]. \label{eq:I-int-uden-fase-rum}
\end{align}
From this we conclude that phase space alone (i.e.~assuming
$|\mathbb{V}(11^{\prime},22^{\prime},33^{\prime})|^2$ to be a
constant) gives a temperature dependence of the form
\begin{align}
I^{\textrm{int}}\propto \e^{-\frac{\TF}{T}} \ T^4 \left[\frac{\Delta
T}{T}\frac{\TF}{T}+\frac{eV}{\kb T}\right]
\quad \textrm{(phase space only)}.
\end{align}
However, as we have seen the three-particle interaction rate has
delicate momentum dependence that needs to be taken into account.
Therefore, to calculate the direct interaction term
$\mathbb{V}(11^{\prime},22^{\prime},33^{\prime})$ we expand the
symmetrized potential $\widetilde{V}_q$ for small $q$, as
\begin{align}
\widetilde{V}_q=V_0
\left[1-\left(\frac{q}{q_0}\right)^2+\mathcal{O}(q^4)\right],
\label{eq:Vq-taylor-series}
\end{align}
where the parameter $q_0\ll\kf$ describes the screening due to the
metallic gates near the quantum wire and $V_0$ is (twice) the $q=0$
Fourier transform of the Coulomb potential cut-off by the screening.
Setting $q_3=q_1\equiv q$ into the three-particle scattering rate
Eq.~(\ref{eq:V-direct-quadratic-sec4}), we obtain
\begin{align}
\mathbb{V}(11^{\prime},22^{\prime},33^{\prime})\propto V_0^2
\left(\frac{\kf}{q_0}\right)^2 q^2
\end{align}
to lowest order in $q$. Inserting this into the
Eq.~(\ref{eq:I-int-uden-fase-rum}) the final result for the current
correction, including both phase space factors and the momentum
dependent scattering rate, becomes
\begin{align}
I^{\textrm{int}}\propto e^{-\frac{\TF}{T}}  \ T^8 V_0^4
\left(\frac{\kf}{q_0}\right)^4 \left[\frac{\Delta
T}{T}\frac{\TF}{T}+\frac{eV}{\kb T}\right].
\end{align}
(Here we noticed that the non-constant three particle scattering
rate gave rise to four extra powers in temperature.) The detailed
calculation given in Appendix \ref{Appendix-B-detailed-calcualtion}
yields a prefactor, and the end result is
\begin{align}
I^{\textrm{int}}=
&\frac{8505}{2048\pi^4}e^{-\frac{\TF}{T}} \frac{e}{\h}\frac{(V_0\kf)^4}{\e_{\textrm{F}}^3}
(L\kf) \bigg(\frac{\kf}{q_0}\bigg)^4 \nonumber \\
&\times \bigg(\frac{T}{\TF}\bigg)^7 \bigg[\frac{\Delta
T}{T}+\frac{eV}{\eF} \bigg]
+\mathcal{O}\left[\bigg(\frac{T}{\TF}\bigg)^8\right].
\label{eq:I-int-resultat}
\end{align}
Combining this result with the zero-order in the interaction terms,
see Eqs.~(\ref{g0}) and (\ref{s0}), we find for the thermopower
and conductance in the low temperature limit,
\begin{align}
S=&\frac{\kb}{e}\frac{\TF}{T}e^{-\TF/T}\left[1
+\frac{L}{\ell_{\textrm{eee}}}\right],
\label{eq:S-result}\\
G=& \frac{2e^2}{h}
   -\frac{2e^2}{h}e^{-\TF/T}\left[1+\frac{L}{\ell_{\textrm{eee}}}\right].
\label{eq:G-result}
\end{align}
Here we introduced the effective length $\ell_{\textrm{eee}}$ by the
relation\cite{sign_of_correction}
\begin{align}\label{elldef}
\ell_{\textrm{eee}}^{-1}=\frac{8505}{2048\pi^3}
\frac{(V_0\kf)^4}{\e_{\textrm{F}}^4} \bigg(\frac{\kf}{q_0}\bigg)^4
\bigg(\frac{T}{\TF}\bigg)^7 \kf,
\end{align}
which may be viewed as a mean free path with respect to
backscattering for a hole near the bottom of the band.

To recapitulate, the temperature dependence $T\propto T^7$ in
Eq.~\eqref{elldef} can be understood in the following way: the
three-particle scattering of a single particle leaves five free
momenta, and since two are taken by energy and momentum conservation
this gives $T^3$. In addition, the interaction, $V_q$, is
proportional to $q^2$, and when squared it gives rise to four more
powers, which results in the $T^7$ dependence.

In the limit of a point-like interaction, $q_0\rightarrow \infty$,
the corrections are zero in agreement with the result of section
\ref{subsec:point-like-int}.

It is known from the Luttinger liquid theory that in the limit of
linear spectrum, which corresponds to $T_F\to\infty$, the
conductance remains finite even if the wire is infinitely long
($L\to\infty$). Therefore it is tempting to speculate that the two
terms in the square brackets of Eq.~(\ref{eq:G-result}) are the
first terms of an expansion in $\lambda=L/\ell_{\textrm{eee}}$ of
some function $F_G(\lambda)$ which saturates at a constant value in
the limit $\lambda\to\infty$. One may also have a similar
speculation generalizing Eq.~(\ref{eq:S-result}) for the
thermopower, $[\dots]\to F_S(L/\ell_{\textrm{eee}})$.

As a final remark, we note that the so-called Mott
formula~\cite{Mott-Book-1936} relating the thermopower to the
low-temperature conductance,
\begin{align}
S=\frac{\pi^2}{3} \frac{\kb}{e} \kb T \frac{1}{G}
\frac{\dd G}{\dd \eF},
\label{mott}
\end{align}
is clearly violated by Eqs.~(\ref{eq:G-result}) and
(\ref{eq:S-result}). This violation could be expected, because the
conventional derivation of the Mott formula (for the non-interacting
case) assumes that the main contribution to the conductance {\it
and} thermopower comes from the states around the Fermi level, in an
energy interval of the order of temperature.\cite{LundeJPCM2005}
However, in the considered case the main contribution to $S$ comes
from the ``deep'' states, even in the zeroth order with respect to
the interaction potential. Correspondingly there is no surprise that
Eqs.~(\ref{eq:S-result}) and (\ref{eq:G-result}) being substituted,
respectively, in the left- and right-hand sides of Eq.~(\ref{mott})
produce a parametrically large mismatch $\sim T_F/T$.

\section{Summary and discussion}

We have calculated the leading interaction correction to the
transport properties of a clean mesoscopic wire adiabatically
connected to the leads, using perturbation theory in the length of
the wire.

For a single-mode wire, the leading interaction corrections turns
out to be given by three-particle scattering processes. This is
because two-particle processes cannot change the current due to
momentum and energy conservation. To calculate the effect of the
three-body processes, we have utilized the Boltzmann equation
formalism, with three-particle scattering events defining the
collision integral. We have identified the leading-order scattering
processes and found that they involve at least one state near the
bottom of the band, i.e. far from the Fermi level. The involvement
of such ``deep'' states results in an exponentially small, $\propto
e^{-T_F/T}$, interaction-induced correction to thermopower and
conductance at low temperatures.

The account for interaction in this paper is performed for relatively
short wires, where perturbation theory in the interaction or
equivalently in the wire length is valid. For longer wires one needs
to find the distribution function by treating the collision integral
in the Boltzmann equation non-perturbatively. It is not clear
whether the relaxation of the distribution function would instead
yield non-exponential corrections to the transport coefficients for
longer wires. However, since the scattering processes that contribute
to the current must involve a particle that changes direction (which
is proven in Appendix \ref{Appendix-A-meso-princip}), one might
speculate that the exponential suppression is valid for all lengths,
as long as electron-electron scattering is the only active relaxation
mechanism.

The question of what the relaxed distribution function looks like
for a mesoscopic wire is an interesting and unsolved problem. Here
we have only given a partial answer for the leading contributions
for a short wire, i.e. to lowest order in the interaction. Further
studies should involve a self-consistent determination of the
distribution function.

Since thermopower is sensitive to the electron distribution function,
it might be a good experimental tool for answering the fundamental
questions regarding the effect of electron-electron collisions.
Indeed refined measurements of thermopower of short 1D quantum wires
have been performed yielding a reasonably good agreement with the
free-electron
theory.\cite{MolenkampPRL1990,MolenkampPRL1992,Appleyard1998} It
remains an open question whether the accuracy of thermopower
measurements is high enough to see the interaction effects in longer
wires.

\section{Acknowledgements}

We acknowledge illuminating discussions with M. Garst, A. Kamenev, M.
Khodas, T. L. Larsen and M. Pustilnik. A.M.L. appreciates and enjoyed
the hospitality of the William I. Fine Theoretical Physics Institute,
University of Minnesota. This work is supported by NSF grants DMR
02-37296 and DMR 04-39026.  \bigskip

\appendix
\section{Scattering processes contributing to the current}\label{Appendix-A-meso-princip}

In this Appendix, we show that \emph{the particle current change
due to electronic scattering if and only if the scattering changes
the number of left and right moving electrons}. In the main text
(see Eq.(\ref{eq:I-1-linear-simplificeret-exact})), this was shown
to first order in the transition rate, but here we show it to
\emph{all orders in the interaction}.

We show it explicitly in the Boltzmann equation framework,
however, suspect it to be a general feature of mesoscopic systems.
Intuitively, the statement means that it is the \emph{number} of
particles that passes though the mesoscopic system that matters
and not their velocity. In contrast to this is e.g.~a long 1D wire
or a bulk metal, where a velocity change of the particles is
enough to change the current.

To show the above statement explicitly, we formally rewrite the
Boltzmann equation (\ref{eq:Boltzmann-eq}) including the boundary
conditions Eq.(\ref{eq:Boundary-conditions}) as
\begin{align}
f_{k}(x)&=f^0_{\textrm{L}}(\e_{k})+\int_0^x\!\!\! \dd x^{\prime}\frac{\mathcal{I}_{kx^{\prime}}[f]}{v_{k}} \ \ \ \textrm{for}\ k>0,
\label{eq:solution-to-BE-genneral-positive-k}\\
f_{k}(x)&=f^0_{\textrm{R}}(\e_{k})+\int_L^x\!\!\! \dd x^{\prime}\frac{\mathcal{I}_{kx^{\prime}}[f]}{v_{k}} \ \ \ \textrm{for}\ k<0.
\label{eq:solution-to-BE-genneral-negative-k}
\end{align}
Note that this is not a closed solution of the Boltzmann equation,
since the distribution function is still contained inside the
collision integral. However, this rewriting enables us to find the
current without finding the distribution function first, i.e.~by
inserting
Eq.~(\ref{eq:solution-to-BE-genneral-positive-k},\ref{eq:solution-to-BE-genneral-negative-k})
into the current definition
\begin{align}
I&=\frac{(-e)}{L}\sum_{\sigma  k}v_{k} f_{k}(x)
\label{eq:current-def}
\end{align}
and obtain (after a few manipulations):
\begin{align}
I=&\frac{(-e)}{L}\sum_{\sigma  k>0}v_{k}
\left[f^0_{\textrm{L}}(\e_{k})-f^0_{\textrm{R}}(\e_{k})\right]\nonumber\\
&-\frac{(-e)}{L} \int_0^L\!\!\! \dd x \sum_{\sigma
k<0}\mathcal{I}_{kx}[f]
+\frac{(-e)}{L} \int_0^x\!\!\! \dd x^{\prime}
\overbrace{\sum_{\sigma
k}\mathcal{I}_{kx^{\prime}}[f]}^{=0}\nonumber\\
\equiv&\  I^{(0)}+I^{(\textrm{int})},
\end{align}
where the $x$-dependent part can be seen to be zero by changing
variables. We note the cancellation of the velocity in the
distribution function Eqs.
(\ref{eq:solution-to-BE-genneral-positive-k}),
(\ref{eq:solution-to-BE-genneral-negative-k}) and the current
definition Eq.~(\ref{eq:current-def}), which is the origin of the
statement we are showing (as in the first order calculation). A
similar cancellation occurs in the Landauer formula thus relating
the transmission to the conductance.  By using the explicit form of
the collision integral Eq.~(\ref{eq:collision-integral-full}) the
current from the interactions is
\begin{widetext}
\begin{align}
I^{(\textrm{int})}=&
\frac{(-e)}{L}\int_0^L\!\!\! \dd x
\!\!\!\!\!\!\sum_{\substack{\sigma_1\sigma_2\sigma_3\\
\sigma_{1^{\prime}}\sigma_{2^{\prime}}\sigma_{3^{\prime}}}}
\sum_{\substack{k_1<0,k_2,k_3\\k_{1^{\prime}}k_{2^{\prime}}k_{3^{\prime}}}}\!\!\!\!
W_{123;1^{\prime}2^{\prime}3^{\prime}}
\left[f_1f_2f_3(1-f_{1^{\prime}})(1-f_{2^{\prime}})(1-f_{3^{\prime}})
-f_{1^{\prime}}f_{2^{\prime}}f_{3^{\prime}}(1-f_1)(1-f_2)(1-f_3)\right].
\end{align}
We can divide the summation over $k$ quantum number into positive
and negative intervals as in the main text (see section
\ref{sec:linear-response-limit}). The essential point is now, that
all terms that have the same number of positive (and negative)
intervals for the primed and unprimed wave numbers $k$ are zero.
In other words, if the number of left and right moving electrons
does not change then the contribution is zero by symmetry of the
transition rate. We show this cancellation in practice by an
example (using the notation of Eq.(\ref{eq:notation-def-++})):
\begin{align}
&\sum_{\textrm{spin}} \sum_{\substack{-+-\\+--}}
W_{123;1^{\prime}2^{\prime}3^{\prime}}
\left[f_1f_2f_3(1-f_{1^{\prime}})(1-f_{2^{\prime}})(1-f_{3^{\prime}})
-f_{1^{\prime}}f_{2^{\prime}}f_{3^{\prime}}(1-f_1)(1-f_2)(1-f_3)\right]\label{eq:simplification-of-sums}\\
=&\sum_{\textrm{spin}} \sum_{\substack{-+-\\-+-}}
W_{123;1^{\prime}2^{\prime}3^{\prime}}
\left[f_1f_2f_3(1-f_{1^{\prime}})(1-f_{2^{\prime}})(1-f_{3^{\prime}})
-f_{1^{\prime}}f_{2^{\prime}}f_{3^{\prime}}(1-f_1)(1-f_2)(1-f_3)\right]\nonumber\\
=&\sum_{\textrm{spin}} \sum_{\substack{-+-\\-+-}}
W_{123;1^{\prime}2^{\prime}3^{\prime}}
f_1f_2f_3(1-f_{1^{\prime}})(1-f_{2^{\prime}})(1-f_{3^{\prime}})
-
\sum_{\textrm{spin}} \sum_{\substack{-+-\\-+-}}
\underbrace{W_{123;1^{\prime}2^{\prime}3^{\prime}}
f_{1^{\prime}}f_{2^{\prime}}f_{3^{\prime}}(1-f_1)(1-f_2)(1-f_3)}_{\textrm{interchange}\
(123)\leftrightarrows(1^{\prime}2^{\prime}3^{\prime})}=0,\nonumber
\end{align}
\end{widetext}
interchanging $1^{\prime}$ and $2^{\prime}$ at the first equality
using
$W_{123;1^{\prime}2^{\prime}3^{\prime}}=W_{123;2^{\prime}1^{\prime}3^{\prime}}$
and interchanging
$(123)\leftrightarrow(1^{\prime}2^{\prime}3^{\prime})$ in the
second term as indicated. Thereby we have shown to all orders that
to change the current by electronic interactions the number of
left and right movers have to change.

The statement is \emph{not limited} to only three particle
scattering and can be show equivalently for pair interaction
including several bands, electron phonon coupling or any other
interaction with the same kind of symmetry under particle
interchange. Furthermore, the statement is still true if the
collision is non-local in space, since that only introduce some
spatial integrals in the collision integral that can be handled
similarly. Note however that the distribution function can be
changed by processes that does not change the number of left and
right movers.

\section{Detailed calculation of the thermopower and conductance correction due to the three particle
scattering}\label{Appendix-B-detailed-calcualtion}

The purpose of this Appendix is to calculate $I^{\textrm{int}}$ in
Eq.~(\ref{eq:I-int-vigtigt})
\begin{align}
I^{\textrm{int}}\simeq&\
3(-e)\sum_{\textrm{spin}}\sum_{\substack{++-\\+--}}
\Delta_{123;1^{\prime}2^{\prime}3^{\prime}}\nonumber\\
&\times\left[\frac{\Delta T}{\kb T^2_{}}
\left\{-\e_3+\e_{2^{\prime}}+\e_{3^{\prime}}-\eF\right\}-\frac{eV}{\kb
T}\right]. \label{eq:I-int-vigtigt-appendix-B}
\end{align}
in the low-temperature limit, $T\ll\TF$, step by step to find the
prefactor given in Eq.~(\ref{eq:I-int-resultat}). As already
mentioned, we preform the calculation with the scattering process
seen in Fig.~\ref{fig:scattering-picture}(a) in mind. Therefore we
use the Fermi functions as given in Eq.~\eqref{f0s} and the matrix
element entering in the scattering rate from
Eq.~(\ref{eq:matrix-element-direct-only}) and
Eq.~(\ref{eq:V-direct-quadratic-sec4}), i.e.~using a quadratic
dispersion.

We preform the summation over all the $k$ in
Eq.~(\ref{eq:I-int-vigtigt-appendix-B}) in the following way:
First of all, we note that due to the momentum and energy
conservation in the interaction process described, the scattering
of $k_3$ to $k_{3^{\prime}}$ has to be from above to below the
Fermi level, i.e.
\begin{align}
\!\!k_3<-\kf < k_{3^{\prime}}\quad \Rightarrow \quad
\theta(-\kf-k_3) \theta(\kf+k_{3^{\prime}}).
\label{eq:fase-rum-restriktion-k3}
\end{align}
This is due to the signs of $k_2$ and $k_{2^{\prime}}$ and can be
understood as a sign of the difference between the curvature of
the dispersion near the bottom of the band and near the Fermi
level. Next, we introduce the momentum transfer around the Fermi
level $q_i\equiv k_{i^{\prime}}-k_i$ for $i=1,3$ and using the
momentum conservation to do the $k_{2^{\prime}}$ summation, we
obtain
\begin{align}
\sum_{\substack{++-\\+--}} (\cdots) \to \sum_{k_1>0,k_2>0,k_3<0}\
\sum_{q_1,q_3} (\cdots)
\end{align}
remembering the constraint $k_{1^{\prime}}=k_1+q_1>0$,
$k_{2^{\prime}}=k_2+q_1+q_3<0$ and $k_{3^{\prime}}=k_3+q_3<0$. The
Fermi factors $f_1^0(1-f^0_{1^{\prime}})$ and
$f_3^0(1-f^0_{3^{\prime}})$ restrict the momentum transfer $q_1$
and $q_3$ to be much smaller then $\kf$ and the $k_1$ and $k_3$ to
be near the Fermi level for the process in mind. Therefore we can
use the Fermi functions $f_1^0(1-f^0_{1^{\prime}})$ to do the
summation over $k_1$. Assuming slow variation of the scattering
rate over a range of $q_1\ll \kf$ at the Fermi level, the $k_1$
summation becomes
\begin{align}
\sum_{k_1>0} \theta(\kf-k_1)\theta(k_1+q_1-\kf)
=\frac{L}{2\pi}q_1\theta(q_1).
\end{align}
Similarly the $k_3$ summation is done using the phase space
constraint in Eq.~(\ref{eq:fase-rum-restriktion-k3})
\begin{align}
\sum_{k_3<0}\theta(-\kf-k_3) \theta(\kf+k_{3^{\prime}})
=\frac{L}{2\pi}q_3\theta(q_3).
\end{align}
We see that since $k_1$ and $k_3$ are restricted to the Fermi
level, we can insert $k_1\simeq \kf$ and $k_3\simeq-\kf$ in the
rest of the integrand. To do the $k_2>0$ summation, we use the
energy conservation contained in the scattering rate. It is
rewritten as (inserting $k_1= \kf$ and $k_3= -\kf$):
\begin{align}
\delta&(\e_{1^{\prime}}+\e_{2^{\prime}}+\e_{3^{\prime}}-\e_{1}-\e_{2}-\e_{3})\simeq
\frac{m}{\h^2} \frac{1}{|q_1+q_3|}\nonumber\\
&\times\delta\left[k_2-\kf\frac{q_1-q_3}{q_1+q_3}-\frac{1}{2}(q_1+q_3)-
\frac{1}{2}\frac{q_1^2+q_3^2}{q_1+q_3} \right].
\label{eq:energy-conservation}
\end{align}

We have now done the summation over $k_1$, $k_2$ and $k_3$ and are
left with the summation over $q_1$ and $q_3$ of the scattering
rate, some Fermi functions and the phase factors described above.
To this end, we introduce $u(q_1,q_3)$ by inserting $k_1= \kf$ and
$k_3=-\kf$ in Eq.~(\ref{eq:V-direct-quadratic-sec4})
\begin{align}
\mathbb{V}(11^{\prime},22^{\prime},33^{\prime})=
\frac{\delta_{\sigma_{1^{\prime}},\sigma_1^{}}
\delta_{\sigma_{2^{\prime}},\sigma_2^{}}
\delta_{\sigma_{3^{\prime}},\sigma_3^{}}}{4L^2 \h^2/m}
\ u(q_1,q_3)
\end{align}
for a general symmetrized interaction
$\widetilde{V}_{q}=V_q+V_{-q}$.

\begin{figure}
\begin{center}
\includegraphics[width=0.35\textwidth,clip=false]{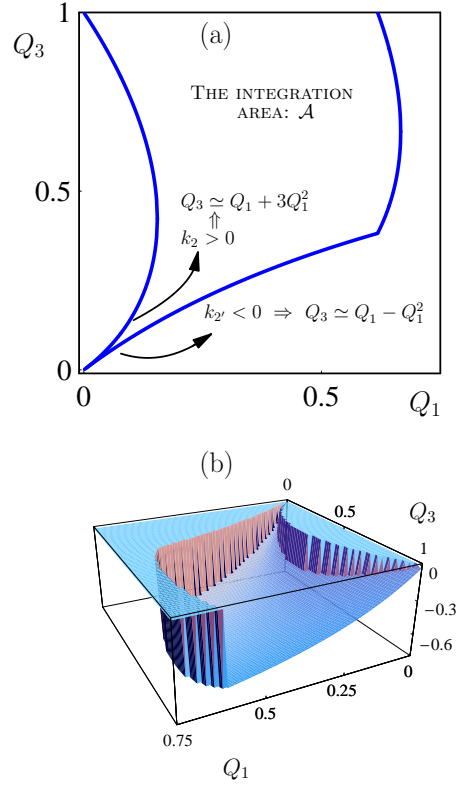}
\caption{(a) The integration region $\mathcal{A}$ for the integral
Eq.~(\ref{eq:three-particle-contribution-to-current-midway-2-dimless})
to calculate the current due to interactions. The two boundaries
for the integration area \emph{close to the origin} stemming from
the signs of $k_2$ and $k_{2^{\prime}}$ are indicated.  (b) The
$\Xi(Q_1,Q_3)=\xi(\kf Q_1,\kf Q_3)/\eF$ function, a dimensionless
version of $\xi(q_1,q_3)$ Eq.~(\ref{eq:xi-def}), important in the
calculation using the method of steepest decent.
\label{fig:int-area}}
\end{center}
\end{figure}

Furthermore, we collect the exponential tales of the Fermi
functions
Eq.~(\ref{eq:Fermi-function-2-2m}-\ref{eq:Fermi-function-3-3m}),
\begin{align}
(1-f^0_{2^{\prime}})f^0_3(1-f^0_{3^{\prime}})=e^{(\e_{2^{\prime}}-\e_3+\e_{3^{\prime}}-\eF)/\kb
T}
\end{align} defining
\begin{align}
&\xi(q_1,q_3)\equiv\e_{2^{\prime}}-\e_3+\e_{3^{\prime}}\label{eq:xi-def} \\
&=\eF\left(\frac{q_1-q_3}{q_3+q_1}\right)^2
-\frac{1}{2}\h \vf (q_1+q_3) \nonumber\\
&+\frac{1}{2}\h \vf \frac{(q_1-q_3)(q_1^2+q_3^2)}{(q_3+q_1)^2}
+\frac{\h^2}{2m}\frac{q_3^2(2q_1^2+2q_1q_3+q_3^2)}{(q_1+q_3)^2},\nonumber
\end{align}
inserting $k_{2^{\prime}}=k_2-q_1-q_3$, $k_2$ from the energy
conservation Eq.~(\ref{eq:energy-conservation}) and $k_1=\kf$ and
$k_3=-\kf$. Therefore we finally get the interacting contribution
to the current in Eq.~(\ref{eq:I-int-vigtigt-appendix-B}) as:
\begin{align}
I^{\textrm{int}}=&
\frac{3(-e)L m^3}{32\pi^4\h^7} \int_0^{\infty} \!\!\!\! \dd q_1
\!\! \int_0^{\infty} \!\!\!\! \dd q_3
\frac{q_1q_3}{q_1+q_3} |u(q_1,q_3)|^2\\
&\times\theta(\kf-k_2)\theta(k_2)\theta(-k_2+q_1+q_3)\theta(\kf-q_3)\nonumber\\
&\times e^{(\xi(q_1,q_3)-\eF)/\kb T}
\bigg[\frac{\Delta T}{\kb T^2_{}}(\xi(q_1,q_3)-\eF)-\frac{eV}{\kb
T} \bigg].\nonumber
\end{align}
Here only the step functions, that restricts the integral are
included. Next we introduce the dimensionless integration
variables $Q_i=q_i/\kf$ for $i=1,3$ and the dimensionless
functions
\begin{align}
U(Q_1,Q_3)&=k_{\textrm{F}}^2\ u(\kf Q_1,\kf Q_3),\\
\Xi(Q_1,Q_3)&=\frac{\xi(\kf Q_1,\kf Q_3)}{\eF}
\end{align}
in the integral:
\begin{align}
I^{\textrm{int}}&=
\frac{3(-e)L m^3}{32\pi^4\h^7\kf}\ e^{-\frac{\TF}{T}}\!\!\!
\int_{\mathcal{A}} \!\!\! \dd Q_1 \dd Q_3
\frac{Q_1Q_3}{Q_1+Q_3} |U(Q_1,Q_3)|^2 \nonumber \\
%
%
\times &e^{\frac{\TF}{T}\Xi(Q_1,Q_3)}
\bigg[\frac{\TF\Delta T}{ T^2_{}}(\Xi(Q_1,Q_3)-1)-\frac{eV}{\kb T}
\bigg],
\label{eq:three-particle-contribution-to-current-midway-2-dimless}
\end{align}
where $\mathcal{A}$ is the integration area shown in
Fig.~\ref{fig:int-area}(a). Note that this expression is valid for
a general interaction $\widetilde{V}_q$ and that it is not
possible to extract a power law in temperature times some integral
by defining new integration variables.

To proceed, we consider the low temperature limit $T/\TF\ll1$ by
using the method of steepest decent: Due to the exponential
function $e^{\frac{\TF}{T}\Xi(Q_1,Q_3)}$ the maximum of
$\Xi(Q_1,Q_3)$ will dominate the integral for $T/\TF\rightarrow0$,
since $\Xi(Q_1,Q_3)\leq 0$. The maxima are $\Xi(0,0)=0$ and
$\Xi(0,1)=0$ and $\Xi(Q_1,Q_3)$ is shown in
Fig.~\ref{fig:int-area}(b). For a decreasing interaction the area
of $Q_1\ll1$ and $Q_3\ll1$ dominates even though the integrand is
zero for $Q_1=Q_3\rightarrow0$. Therefore we expand around the
maximum $(Q_1,Q_3)=(0,0)$ to get the lowest order result in
$T/\TF$. In view of the integration region
Fig.~\ref{fig:int-area}(a), we use $Q_3=Q_1\equiv Q$ in the
integral
Eq.(\ref{eq:three-particle-contribution-to-current-midway-2-dimless})
and thereby do the $Q_3$ integral using the approximate limits
seen in Fig.~\ref{fig:int-area}(a), i.e.
\begin{align}
\int_{Q_1-Q_1^2}^{Q_1+3Q_1^2} \!\!\!\!\!\!\! 1 \dd Q_3=4Q_1^2.
\end{align}
To model the symmetrized potential $\widetilde{V}_q$ for small
$q$, we include the deviation from a constant as described in
Eq.~(\ref{eq:Vq-taylor-series}). This gives
\begin{align}
\hspace{-2mm}\frac{Q_1Q_3}{Q_1+Q_3}
|U(Q_1,Q_3)|^2\bigg|_{Q_3=Q_1\equiv Q}
\!\!\!\!\rightarrow V_0^4 \left(\frac{\kf}{q_0}\right)^4
\frac{9}{2} Q^5
\end{align}
to lowest order in $Q$. In the exponential we keep $\Xi$ to lowest
order in $Q$, i.e.
\begin{align}
e^{\frac{\TF}{T}\Xi(Q_1,Q_3)}\rightarrow e^{-2Q\frac{\TF}{T}}.
\end{align}
So using the lowest order in $Q$ in the integrand (leading to
lowest order in $T$) the interacting contribution to the current
is:
\begin{align}
I^{\textrm{int}}=&
\frac{3(-e)L m^3}{32\pi^4\h^7\kf}\ e^{-\frac{\TF}{T}}\!\!\!
\int_{0}^{\infty} \!\!\! \dd Q \
V_0^4 \left(\frac{\kf}{q_0}\right)^4 \frac{9}{2} Q^5 \nonumber \\
&\times  4Q^2 e^{-2Q\frac{\TF}{T}}
\bigg[\frac{\TF\Delta T}{T^2_{}}(0-1)-\frac{eV}{\kb T} \bigg]\\
\simeq& \frac{8505}{2048\pi^4}e^{-\frac{\TF}{T}} \frac{e}{\h}\frac{(V_0\kf)^4}{\e_{\textrm{F}}^3}\nonumber \\
&\times (L\kf) \bigg(\frac{\kf}{q_0}\bigg)^4
\bigg(\frac{T}{\TF}\bigg)^7
\bigg[\frac{\Delta T}{T}+\frac{eV}{\eF} \bigg]
\end{align}
to lowest order in temperature. This is the result stated in the
text in Eq.~(\ref{eq:I-int-resultat}).


\begin{thebibliography}{10}
\bibitem{coulomb} We assume that electrons in the channel produce some
  image charges in a distant gate. As the result, the Coulomb
  interaction is cut off at some distance $\sim 2\pi/q_0$ large
  compared to the Fermi wavelength.
\bibitem{sign_of_correction}
Note here that the sign of the conductance correction is negative
even though there is a positive sign in
Eq.(\ref{eq:I-int-resultat}). This is due to the way the bias is
defined the left contact having the higher chemical potential.
\end{thebibliography}
\end{document}